\documentclass[sigconf,nonacm]{acmart}
\usepackage{mathtools}
\usepackage{fontawesome5}
\usepackage{tabularx}
\usepackage{multirow}
\usepackage{colortbl}

\usepackage{enumitem}

\AtBeginDocument{%
  }

\setcopyright{none}
\makeatletter
\newcommand{\preprintcodelink}{%
  \par\vspace{8pt}\noindent
  {\fontsize{10}{12}\selectfont
  \href{https://github.com/KennyCaty/SCI-Agent}{%
    \textcolor{black}{\faGithub}\enspace\textcolor[HTML]{000080}{\texttt{https://github.com/KennyCaty/SCI-Agent}}}\par}}
\patchcmd{\@typeset@author@bx}
  {\@currentaffiliation\par}
  {\@currentaffiliation\par\preprintcodelink}
  {}{\PackageError{preprint}{Could not insert the code link}{Check the acmart version.}}
\newcommand{\preprintemailseparator}{\par\vspace{5pt}\gdef\preprintemailseparator{\hspace{2em}}}
\patchcmd{\@mkauthors@iii}
  {\par\bgroup}
  {\preprintemailseparator\bgroup\fontsize{10}{12}\selectfont\ttfamily}
  {}{\PackageError{preprint}{Could not apply inline email layout}{Check the acmart version.}}
\makeatother

\begin{document}

\title{When Connectivity Misleads: Guiding Agentic Search with Contrastive Evidence Paths}

\author{Junchen Li, Liang Xu, Rongzheng Wang, Qizhi Chen, Junhao Wen, Shuang Liang}
\correspondingauthor
\affiliation{%
  \institution{University of Electronic Science and Technology of China}
  \city{Chengdu}
  \country{China}}
\email{junchenli@std.uestc.edu.cn}
\email{shuangliang@uestc.edu.cn}
\renewcommand{\shortauthors}{Li et al.}

\begin{abstract}
Search agents can use cross-document graphs to discover sources through multi-hop connections and choose subsequent actions based on candidate rankings and relationships within a limited observation window. However, documents may rank highly because their numerous connections make them easier to reach, without necessarily providing more valuable evidence. Graph diffusion methods such as Personalized PageRank (PPR) aggregate propagation contributions from multiple sources, allowing highly connected distractor nodes to crowd the observation window. Flat node scores also obscure cross-document evidence trails, limiting continued exploration. We propose Contrastive Evidence Paths (CEP), which scores candidates by their strongest individual paths on a query-conditioned document graph and calibrates these scores against the same graph's background visitation distribution to mitigate structural bias. CEP ranks candidates by the calibrated scores and retains the corresponding source paths. Building on CEP, we develop Structured Corpus Interaction (SCI), which organizes candidates and their connecting paths into a structured map, enabling agents to verify source documents and pursue missing information through expansion from anchor documents. 
Experiments on BrowseComp-Plus and TREC-COVID show that SCI achieves state-of-the-art answer accuracy and nDCG@10 with the lowest generation cost. Compared with the strongest baselines by task performance, maximum reductions in tool calls and total generation costs exceed 50\% and 40\%, respectively. Further analysis confirms that SCI with CEP suppresses ranking inflation induced by structural connections and improves gold recall and exploration efficiency.
\end{abstract}



\maketitle

\section{Introduction}
To address complex information needs in question answering, fact checking, and deep research on the Web, 
retrieval-augmented generation (RAG)~\cite{RAG} is evolving from one-shot retrieval toward search agents that interact continuously with a corpus. Methods such as Direct Corpus Interaction (DCI)~\cite{dci,drdci,rise,rarg} allow agents to search, read, and programmatically process source documents, adapting their actions to newly discovered clues. 
However, when evidence spans documents with weak direct similarity to the query, lexical or semantic similarity matching alone may miss connections through intermediate entities.
Recent work therefore combines cross-document graphs with interactive corpus access~\cite{atlasnav}, allowing agents to discover sources across multiple hops and continue exploring through precomputed document connections.

\begin{figure}[!t]
\centering
\captionsetup{skip=4pt}
\includegraphics[width=\columnwidth]{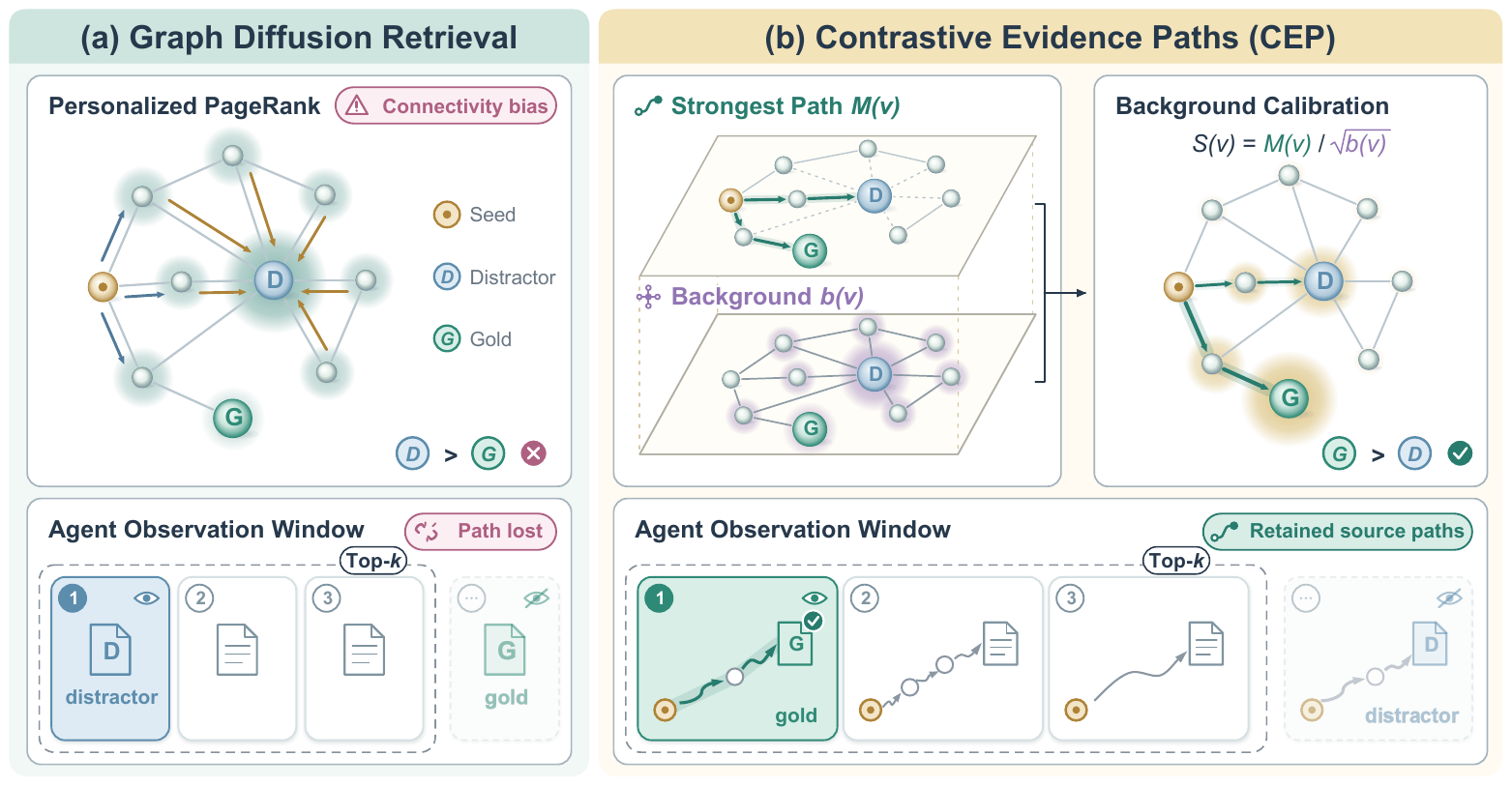}
\caption{Illustrative comparison of graph diffusion retrieval and contrastive evidence paths retrieval.}
\label{fig:motivation}
\Description{Conceptual comparison on the same document graph. Graph diffusion prioritizes a highly connected distractor and exposes a flat node ranking. CEP combines strongest path support with a background visitation distribution to calibrate candidate priorities and retains source paths in the agent observation window.}
\end{figure}

Although graphs broaden evidence discovery, their utility for search agents depends on which candidates are prioritized and how those candidates are presented. 
At each turn, agents typically observe only a limited set of candidate previews. 
In exploratory tasks such as deep research on the Web, candidate rankings and the connections exposed in retrieval results guide agents in selecting sources to verify and following clues across sources to pursue further evidence.
Several graph retrieval methods employ graph diffusion algorithms, such as Personalized PageRank (PPR), to prioritize candidate documents based on visitation distributions over graph nodes~\cite{HippoRAG,HippoRAG2,linearRAG}. 
However, using such results to guide agent exploration raises two challenges, as shown in Figure~\ref{fig:motivation} (a).

\textbf{Challenge 1: Connectivity Induced Ranking Bias.}
Graph diffusion scores can be biased by graph connectivity~\cite{normalization}.
PPR aggregates propagation contributions along multiple paths, which can give highly connected distractors high scores. Our controlled experiments on the document graph demonstrate this effect: with document content, queries, and initial retrieval seeds fixed, adding connections to selected distractor documents can substantially improve their PPR ranks, even bringing some into the top 10. Thus, propagation advantages arising from connectivity can prioritize distractors within an agent's limited observation window.

\textbf{Challenge 2: Path Blindness in Exploration.}
Even with graph retrieval, flat node scores do not directly expose the source paths connecting candidate documents to retrieval seeds.
Corpus interaction agents retrieve candidate documents into a local workspace to narrow the search space, but must still rely on clues within a limited observation window to decide what to search for and which documents to read next. Without explicit paths and mechanisms for directed expansion, gold documents in the workspace may remain unread. 
The key is to integrate graph retrieval with corpus interaction, allowing retrieved candidates and their source paths to guide agents in document verification and continued exploration.

To address these challenges, we propose Contrastive Evidence Paths (CEP), a path retrieval algorithm combining strongest path support with background calibration (Figure~\ref{fig:motivation}, right). CEP uses question-conditioned transitions on a document graph connected through shared entities. It scores each candidate by the maximum path score from query-relevant seeds and retains the corresponding source path. This scoring rule evaluates candidates by their strongest individual path, avoiding score inflation from accumulating contributions across multiple paths. To mitigate ranking bias arising from connectivity, CEP estimates a background visitation distribution through random walks with uniform restart and uses it to calibrate candidate scores. This requires easily reached documents to have stronger query-specific path support to rank highly.

Building on the candidate rankings and source paths produced by CEP, we develop Structured Corpus Interaction (SCI) to support agents' ongoing evidence exploration. Through the global retrieval tool \textbf{\texttt{graphsearch}}, SCI collects the full text of candidate documents and documents along their strongest paths into a workspace. A structured map presents top-ranked candidates, source paths, shared entities, and text previews to guide agents' next actions. Agents can use \textbf{\texttt{read}} to read documents, use \textbf{\texttt{bash}} to search and verify source text within the workspace, or select discovered documents as anchors and use \textbf{\texttt{graphexpand}} to explore related sources that address information gaps.

We evaluate SCI on BrowseComp-Plus~\cite{bcp} for complex question answering and TREC-COVID~\cite{treccovidbeir} for scientific document retrieval. Across three backbones on BrowseComp-Plus, SCI achieves the highest accuracy, fewest tool calls, and lowest total generation cost among the evaluated methods. Compared with the most accurate baseline for each backbone, SCI reduces tool calls by 14.9\%--54.0\% and total generation costs by 8.4\%--38.6\%. SCI also achieves the highest nDCG@10 and lowest total generation cost on TREC-COVID. Ablations and trajectory analysis show that the map and anchor expansion bring new gold documents into the workspace and expose previously undisplayed gold documents for subsequent reading. Mechanism analysis shows that CEP reduces the preference for highly connected nodes, improves gold document rankings, and suppresses distractor ranking gains from additional connections.

Our main contributions are as follows:
\begin{itemize}[noitemsep,topsep=0pt,partopsep=0pt,parsep=0pt]
\item We identify a mismatch between graph retrieval rankings and evidence needs: propagation advantages arising from connectivity can give distractor documents high rankings.
\item We propose Contrastive Evidence Paths (CEP), which combines strongest path support with background visitation calibration to mitigate connectivity bias and retain source paths for candidate documents.
\item We develop Structured Corpus Interaction (SCI), integrating CEP retrieval with a structured map and anchor expansion to support source verification and continued evidence exploration.
\item We evaluate SCI on BrowseComp-Plus and TREC-COVID, demonstrating higher answer accuracy and document ranking quality with fewer tool calls and lower generation costs. Further analyses demonstrate the benefits of path support, background calibration, and structured corpus interaction.
\end{itemize}

\section{Related Work}
Our work connects two lines of research: corpus interaction for search agents and graph structures for evidence retrieval.

\textbf{Corpus Interaction for Search Agents.} Search agents gather and verify evidence through iterative retrieval and reading. DCI~\cite{dci} enables agents to search and read raw corpora through terminal commands, but repeated scans of the entire corpus increase latency and interaction costs as corpora grow. DR-DCI~\cite{drdci} and RISE~\cite{rise} expose retrieval as an action that agents can invoke to progressively collect relevant documents in a local workspace for detailed search and verification. RARG++~\cite{rarg} further uses relevance signals to prioritize document traversal, select entry passages, and rerank search matches. AtlasNav~\cite{atlasnav} organizes document graphs with multiple views into hierarchical regions through Leiden community detection~\cite{Leiden}, enabling agents to navigate the corpus structure. However, its candidate scoring relies on retrieval rank fusion without directly incorporating path support from query seeds to candidate documents. SCI combines candidate ranking with source paths to guide retrieval, reading, and expansion from anchors.

\textbf{Graph Structures for Retrieval.} Graph structures connect entities and organize relations across documents, enabling retrieval of sources beyond direct text matching~\cite{LightRAG,neuropath,HippoRAG2}. GraphRAG~\cite{graphrag} constructs entity graphs and community summaries for question answering over an entire corpus. HippoRAG~\cite{HippoRAG} and LinearRAG~\cite{linearRAG} use Personalized PageRank (PPR) to propagate query signals and retrieve relevant information through graph connections. Such propagation aggregates contributions from multiple sources into node scores, allowing document connectivity to influence retrieval priorities. PathRAG~\cite{PathRAG} selects relational paths through flow propagation and pruning for answer generation. Its propagation depends on connectivity, potentially favoring highly connected nodes. We study how to exploit relations across documents while mitigating ranking bias arising from connectivity itself.

\begin{figure*}[t]
\centering
\captionsetup{skip=4pt}
\includegraphics[width=\textwidth]{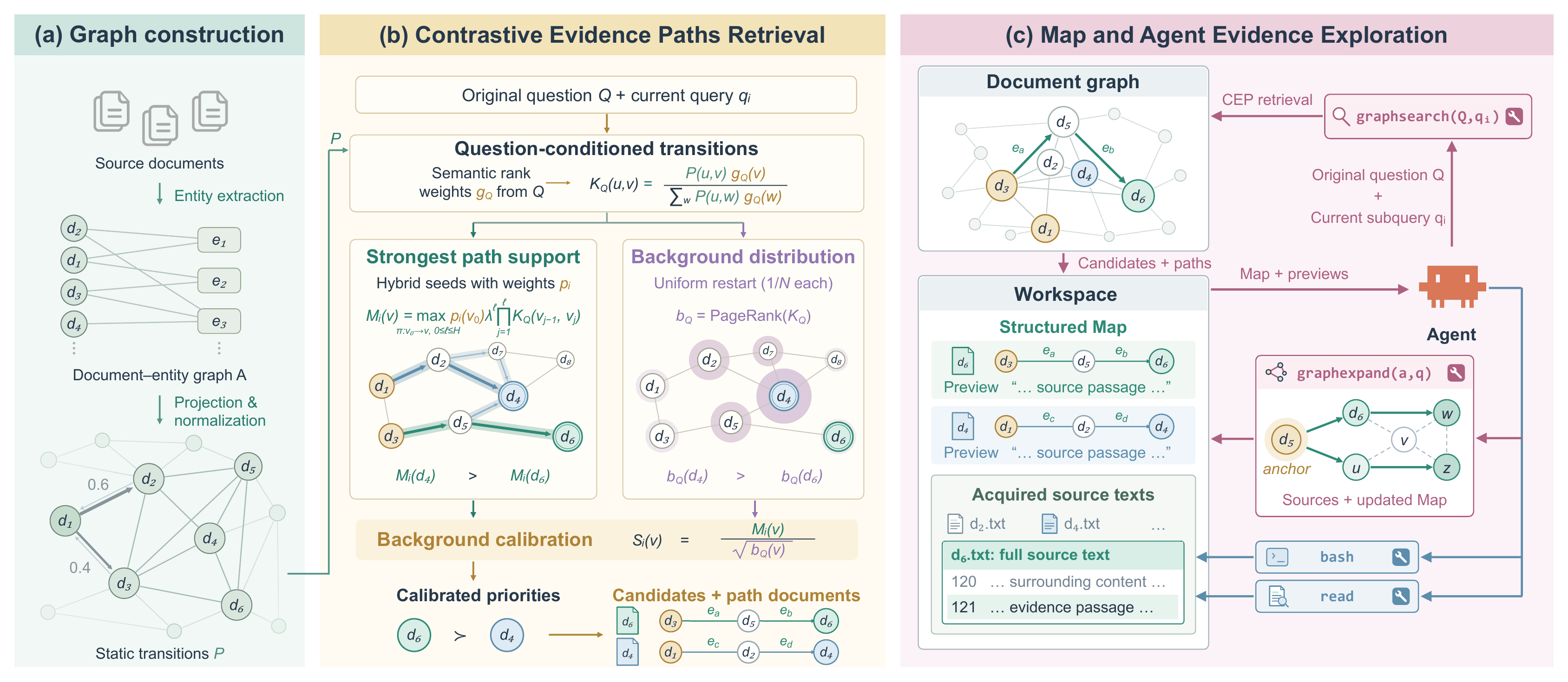}
\caption{Overview of SCI. (a) Shared entities connect source documents into a document graph with static transition probabilities. (b) CEP adjusts transition probabilities using the original question, computes strongest path support $M_i(v)$ from query-relevant starting points, and calibrates candidate scores using the background visitation distribution $b_Q(v)$. It collects documents and builds a map. (c) The agent uses the map and previews to choose actions: verify collected source text with \textbf{\texttt{read}} and \textbf{\texttt{bash}}, expand from an anchor to address an information gap with \textbf{\texttt{graphexpand(a,q)}}, or initiate global retrieval with \textbf{\texttt{graphsearch}}.}
\label{fig:method}
\Description{Shared entities connect documents into a graph. The original question adjusts transition probabilities, while the original question and current query jointly select starting points. CEP calibrates strongest path support using background visits and collects documents into a workspace with a navigation map. The agent explores through reading, command-line checks, global search, and anchor expansion.}
\end{figure*}

\section{Problem Formulation and Preliminaries}
\subsection{Corpus Interaction Setting}
We use a search agent based on Direct Corpus Interaction (DCI)~\cite{dci}. Given a document corpus and an input query, the agent searches for and verifies evidence through basic tools: \textbf{\texttt{bash(command)}} executes commands to search document content, and \textbf{\texttt{read(path, offset, limit)}} reads a specified span from a document. Repeated searches over large corpora can be costly, so retrieval tools collect the full text of relevant documents into a local workspace for search and reading. The agent uses tool outputs to decide subsequent actions, accumulate evidence, and complete the task.

Documents stored in the workspace are not all presented in the agent's context at once. Each tool call exposes only limited candidate information or previews. We therefore focus on how to select and present candidate documents to support continued evidence exploration within a limited observation window.

\subsection{Document Graph and Path Notation}\label{sec:background}
We represent relations among documents as a weighted graph $G=(\mathcal V,\mathcal L)$, where $\mathcal V$ is the set of document nodes and $\mathcal L$ is the set of edges representing relations among documents. For a retrieval operation, let $p(v)$ denote the starting weight of document $v$ and $K(u,v)$ the transition probability used during retrieval. A path $\pi=(v_0,\ldots,v_\ell)$ is a sequence of connections from starting document $v_0$ to candidate document $v_\ell$ through $\ell$ transitions.

Given the starting weights and transition probabilities, the base weight of a path is
\[
w(\pi)=p(v_0)\prod_{j=1}^{\ell}K(v_{j-1},v_j),
\]
where the path weight is $p(v_0)$ when $\ell=0$. This notation describes the paths connecting retrieval starting points to candidate documents. The specific transition probabilities, path aggregation, and candidate scoring are defined in the method section.

\section{Method}
Figure~\ref{fig:method} provides an overview of SCI. Offline, we connect documents through shared entities and compute static transition probabilities (Section~\ref{sec:graph}). During retrieval, CEP adjusts these probabilities using the original question, computes strongest path support and a background visitation distribution, and calibrates candidate priorities. It then collects candidates and documents along their paths (Section~\ref{sec:cep}). A map combines navigation links and source text previews to guide reading, command-line verification, expansion from anchors, and further global search (Section~\ref{sec:interaction}). Retrieval results update the workspace to support subsequent decisions.

\subsection{Document Graph Construction}\label{sec:graph}
Given a corpus $\mathcal V$ of $N$ documents, we use a pretrained information extraction model to identify entity mentions, types, and confidence scores. Mentions with the same type and normalized text are treated as the same entity. Let $\operatorname{df}(e)$ denote the number of documents containing entity $e$. We retain entities satisfying $2\leq\operatorname{df}(e)\leq\tau$ to form the entity set $\mathcal E$, where $\tau$ is the document frequency limit. This selects entities shared across documents and controls the number of connections introduced by frequent entities. Extraction settings and thresholds appear in Appendix~\ref{app:experiments}.

Documents $\mathcal V$ and entities $\mathcal E$ form a bipartite graph, with each document connected to the entities it contains. We compute association weights from mention counts and confidence scores, then normalize per entity:
\begin{equation}
\widetilde A_{ve}=\log(1+c_{ve})\gamma_{ve},
\qquad
A_{ve}=\frac{\widetilde A_{ve}}
{\sum_{u\in\mathcal V}\widetilde A_{ue}},
\label{eq:1}
\end{equation}
where $c_{ve}$ is the number of mentions of entity $e$ in document $v$, $\gamma_{ve}$ is their maximum confidence score, and $\widetilde A_{ve}$ and $A_{ve}$ are the association weights before and after normalization. The logarithm reduces the marginal contribution of repeated mentions.

We project the bipartite graph into a document graph, defining edge weights through shared entities:
\begin{equation}
W_{uv}=
\begin{cases}
\displaystyle\sum_{e\in\mathcal E}A_{ue}A_{ve},&u\neq v,\\
0,&u=v.
\end{cases}
\label{eq:2}
\end{equation}
With equal mention counts and confidence scores, the normalization in Eq.~\eqref{eq:1} reduces the contribution of frequent entities to individual edge weights. Row normalization of the edge weights yields the static transition matrix $P$:
\begin{equation}
P(u,v)=
\begin{cases}
\displaystyle\frac{W_{uv}}{\sum_{w\in\mathcal V}W_{uw}},&\displaystyle\sum_{w\in\mathcal V}W_{uw}>0,\\
0,&\text{otherwise}.
\end{cases}
\label{eq:3}
\end{equation}
For a document $u$ with neighbors, $P(u,v)$ is the probability of moving from $u$ to $v$.
This matrix is constructed offline and provides the basis for adjusting transition probabilities using the question.

\subsection{Contrastive Evidence Paths Retrieval}\label{sec:cep}
In this subsection, we describe how CEP ranks candidate documents and collects them with their source paths into the workspace.

\textbf{Starting Points and Question-Conditioned Transitions}
We use hybrid retrieval to select starting points, combining semantic relevance to the original question $Q$ with lexical matching to $q_i$, the $i$-th query in the current retrieval call. For semantic matching, we divide each document $v$ into a set of text segments $\mathcal S(v)$ and encode them offline with a semantic encoder. The document's semantic score is the maximum segment score:
\begin{equation}
d_Q(v)=\max_{x\in\mathcal S(v)}
\langle\mathbf z_Q,\mathbf z_x\rangle,
\label{eq:4}
\end{equation}
where $\mathbf z_Q$ and $\mathbf z_x$ are normalized question and segment embeddings, and $\langle\cdot,\cdot\rangle$ denotes the inner product. Let $r_Q(v)$ denote the document's rank in descending order of $d_Q(v)$.

For query $q_i$, document-level BM25~\cite{bm25} provides the lexical rank $r_i^{\mathrm{lex}}(v)$. We combine it with the semantic rank using reciprocal rank fusion (RRF)~\cite{rrf}:
\begin{equation}
f_i(v)=\frac{1}{k+r_Q(v)}
+\frac{1}{k+r_i^{\mathrm{lex}}(v)},
\label{eq:5}
\end{equation}
where $f_i(v)$ is the fused score, $k$ is a rank offset, and all ranks start at 1. Lexical ranks cover documents with positive BM25 scores. For other documents, the second term in Eq.~\eqref{eq:5} is zero. We select the top $R$ documents by $f_i(v)$ as starting points, denoted $v_{i,1},\ldots,v_{i,R}$, and assign starting weights proportional to their reciprocal ranks:
\begin{equation}
p_i(v_{i,j})=\frac{1/j}{\sum_{a=1}^{R}1/a},\qquad j=1,\ldots,R.
\label{eq:starting-weights}
\end{equation}

To guide path search toward documents relevant to the original question, we adjust the static transition matrix $P$ using semantic ranks. Reweighting destination documents and normalizing each row yields the question-conditioned transition matrix $K_Q$:
\begin{equation}
\begin{aligned}
g_Q(v)&=(k+r_Q(v))^{-1/2},\\[3pt]
K_Q(u,v)&=\frac{P(u,v)g_Q(v)}
{\sum_{w\in\mathcal V}P(u,w)g_Q(w)},
\end{aligned}
\label{eq:6}
\end{equation}
where $g_Q(v)$ is the semantic weight of destination document $v$, and $K_Q(u,v)$ is the adjusted probability of moving from $u$ to $v$. For equal static transition probabilities, neighbors with higher semantic ranks receive higher transition probabilities. Starting weights $p_i$ depend on both $Q$ and $q_i$, while $K_Q$ depends on $Q$.

\textbf{Path Scores with Background Calibration}
We score candidate documents by their strongest path support and use a background visitation distribution to calibrate scores for structural effects. A path's weight combines its starting weight, transition probabilities, and a hop discount. For each candidate document $v$, we take the maximum score over all paths of at most $H$ hops from any retrieval starting point to $v$, defining its strongest path support as:
\begin{equation}
M_i(v)=
\max_{\substack{\pi=(v_0,\ldots,v_\ell=v)\\0\leq\ell\leq H}}
\Bigl[
p_i(v_0)\lambda^\ell
\prod_{j=1}^{\ell}K_Q(v_{j-1},v_j)
\Bigr],
\label{eq:7}
\end{equation}
where $\pi$ is a path from retrieval starting point $v_0$ to candidate document $v$, $\ell$ is its hop count, and $0<\lambda<1$ is the per-hop discount. Zero-hop paths have weight $p_i(v)$, allowing directly retrieved documents to participate in scoring.

To measure general accessibility under $K_Q$, we estimate the background visitation distribution $b_Q$ using PageRank~\cite{pagerank} with uniform restart over all documents. Starting from $b_Q^{(0)}(v)=1/N$, we iterate:
\begin{equation}
b_Q^{(t+1)}(v)=\frac{\alpha}{N}+(1-\alpha)\Bigl[\sum_{u\in\mathcal V}K_Q(u,v)b_Q^{(t)}(u)+\frac{1}{N}\sum_{u\in\mathcal Z}b_Q^{(t)}(u)\Bigr],
\label{eq:background-iteration}
\end{equation}
where $0<\alpha<1$ is the restart probability, $t$ is the iteration index, and $\mathcal Z$ is the set of isolated documents. The final term redistributes their probability uniformly over all documents. After $T_b$ iterations, we use $b_Q=b_Q^{(T_b)}$ as the background visitation distribution.

We normalize strongest path support by the square root of the background visitation probability to obtain the calibrated score:
\begin{equation}
S_i(v)=\frac{M_i(v)}{\sqrt{b_Q(v)}}.
\label{eq:8}
\end{equation}
where the square root moderates the effect of background probability differences on the calibration factor. Documents with higher background visitation probabilities require stronger path support to outrank other candidates.

\textbf{Strongest Path Computation and Backtracking}
We compute strongest path support in Eq.~\eqref{eq:7} using max-product dynamic programming. Let $m_i^{(t)}(v)$ denote the strongest path support at $v$ within $t$ hops. We initialize $m_i^{(0)}(v)=p_i(v)$ and update:
\begin{equation}
m_i^{(t+1)}(v)=\max\Bigl\{
p_i(v),
\lambda\max_{\mathclap{u:P(u,v)>0}}K_Q(u,v)m_i^{(t)}(u)
\Bigr\}.
\label{eq:9}
\end{equation}
The inner maximum is defined as zero when the predecessor set is empty. Each iteration retains the larger of the zero-hop path weight and the weights of paths extended by one edge from a predecessor. After $H$ iterations, we obtain $M_i(v)=m_i^{(H)}(v)$ and compute candidate scores using Eq.~\eqref{eq:8}. This recurrence is a max-product form of generalized Bellman-Ford~\cite{nbfnet}. In practice, we use an equivalent edgewise background calibration that incorporates the calibration in Eq.~\eqref{eq:8} into the path recurrence, as derived in Appendix~\ref{app:equivalence}.

By recording the maximizing choice at each iteration, we recover a candidate's strongest path by tracing predecessors back to a zero-hop start. The document-entity associations in Section~\ref{sec:graph} identify the shared entities between adjacent documents on the path.

Each iteration of background estimation and strongest path computation traverses the documents and their nonzero transitions, taking $O(N+J)$ time, where $N$ is the number of documents and $J$ is the number of nonzero directed transitions. For a retrieval call with $m$ queries, background estimation runs for $T_b$ iterations and strongest path computation runs for $H$ iterations per query, giving a total time complexity of $O((T_b+mH)(N+J))$.

\textbf{Collecting Candidates and Path Documents}
For each query, we select the top $B$ documents by $S_i$ as the candidate set $\mathcal C_i$ and recover each candidate's strongest path $\pi_i^\star(v)$. We add the full text of candidates and documents on their strongest paths to the workspace for the agent to read and verify. Let $\mathcal W$ denote the existing workspace documents. The updated set is
\begin{equation}
\mathcal W'=\mathcal W\cup
\bigcup_{i=1}^{m}\bigcup_{v\in\mathcal C_i}
V(\pi_i^\star(v)),
\label{eq:10}
\end{equation}
where $V(\pi)$ is the set of documents on path $\pi$. All documents are deduplicated by document identifier.

\subsection{Map and Agent Evidence Exploration}\label{sec:interaction}
SCI equips the agent with two retrieval tools: \textbf{\texttt{graphsearch}} for global retrieval and \textbf{\texttt{graphexpand}} for expansion from anchors, alongside the basic tools \textbf{\texttt{bash}} and \textbf{\texttt{read}}.

\textbf{Map-Guided Evidence Exploration}
The agent submits the original question and complementary queries through \textbf{\texttt{graphsearch}}. The tool uses CEP to select $B$ candidates for each query and collects these documents and those along their strongest paths into the workspace. It also builds a map for the top 10 candidates for each query. The map presents source previews of 240 characters and source paths, with shared entities linking adjacent documents. The agent can use \textbf{\texttt{read}} to inspect documents and \textbf{\texttt{bash}} commands, such as \texttt{rg}, to search and verify source text in the workspace.

The agent can also select a document in the map as an anchor and use \textbf{\texttt{graphexpand(a,q)}} to explore related sources, where $a$ is the anchor document and $q$ describes the missing information. Expansion can add previously unacquired source documents and paths to the workspace and update the map. It may also bring documents already in the workspace but previously absent from the map back into view.

\textbf{Anchor-Guided Local Expansion}
To control the computational cost of local expansion, we use beam search based on CEP to explore candidate documents near the anchor. Given anchor document $a$ and expansion query $q$, we compute the transition matrix $K_q$ and background visitation distribution $b_q$ using $q$, as in Section~\ref{sec:cep}. With $a$ as the sole starting point and its weight set to 1, the calibrated score of a path $\pi=(v_0,\ldots,v_\ell)$ is
\begin{equation}
L_{a,q}(\pi)=
\lambda^\ell\prod_{j=1}^{\ell}K_q(v_{j-1},v_j)/\sqrt{b_q(v_\ell)},
\quad v_0=a,\; 1\leq\ell\leq2.
\label{eq:12}
\end{equation}
Search proceeds from the anchor for at most two hops, excluding paths that revisit a document. At each depth, we retain the paths with the highest $L_{a,q}$ scores for further expansion. We pool the documents reached at all depths and rank them by their highest path score found during search. Table~\ref{tab:config} lists search budgets.

\section{Experiments}
\subsection{Experimental Setup}\label{sec:setup}
In this subsection, we describe the datasets, baselines, and implementation details.

\textbf{Datasets.}
We evaluate SCI on BrowseComp-Plus~\cite{bcp}, a deep research benchmark, and TREC-COVID~\cite{treccovidbeir} from BEIR~\cite{BEIR} for scientific literature retrieval. BrowseComp-Plus contains 100,195 Web documents and 830 queries. TREC-COVID contains 171,332 documents, 50 queries, and graded relevance judgments.
Additionally, to evaluate how increasing corpus size affects SCI, we sample 900,000 documents from FineWeb-Edu~\cite{finewebedu} and add them to the original BrowseComp-Plus corpus for corpus scaling experiments. We report accuracy on BrowseComp-Plus and nDCG@10 on TREC-COVID, along with average turns and tool calls per query and total generation cost for agent methods.

\textbf{Models and baselines.}
On BrowseComp-Plus, we compare two groups of baselines. \textit{Conventional retrieval agents} include Retrieval-Agent released with RISE~\cite{rise} and NVIDIA's NeMo Agent~\cite{nemoagent}, which iteratively retrieves and ranks documents. \textit{Corpus interaction agents} include DCI~\cite{dci}, DR-DCI~\cite{drdci}, RISE~\cite{rise}, RARG++~\cite{rarg}, and AtlasNav~\cite{atlasnav}. We evaluate SCI and all these baselines with GPT-6-Luna~\cite{gpt6}, DeepSeek-V4.1-Flash~\cite{deepseek-v4.1-flash}, and Qwen3.8-Flash~\cite{qwen38flash}. On TREC-COVID, we evaluate SCI and the same corpus interaction agents with GPT-6-Luna. For RISE, we use the original paper's RISE-BM25 variant~\cite{rise}, avoiding the substantial cost of offline table of contents generation for the new corpus. Appendix~\ref{app:supplementary} reports additional comparisons of agents with smaller models and comparisons with graph-based RAG methods.

\textbf{Implementation details.}
SCI extracts entities with GLiNER2.5-small~\cite{gliner} and combines Qwen3-Embedding-0.6B~\cite{qwenembed} with BM25 to identify retrieval starting points. The maximum path length is $H=3$ for \textbf{\texttt{graphsearch}}. Following RISE~\cite{rise}, we set the candidate budget to $B=1000$ per query and display the top 10 candidates per query in the map. All agent methods use at most 100 turns and a time budget of 3,600 seconds per question. Full SCI parameter settings are provided in Appendix~\ref{app:experiments}.

\subsection{Main Results}

We report results on SCI's effectiveness, efficiency, and scalability.

\textbf{BrowseComp-Plus question answering.}
As shown in Table~\ref{tab:bcp}, SCI achieves accuracies of 91.08\%, 95.30\%, and 94.94\% with GPT-6-Luna, DeepSeek-V4.1-Flash, and Qwen3.8-Flash, respectively, achieving the highest accuracy, fewest tool calls, and lowest generation cost. 
Compared with the most accurate baseline for each backbone, SCI reduces tool calls by 14.9\%--54.0\% and total generation costs by 8.4\%--38.6\%. With DeepSeek-V4.1-Flash, SCI outperforms RARG++, the strongest baseline, in accuracy while reducing both turns and tool calls by roughly half. Compared with AtlasNav, which also uses graph structures, SCI requires a similar number of turns but roughly half as many tool calls. These results show that SCI uses graph structures more effectively to support evidence exploration, improving answer accuracy while reducing tool calls.

Compared with conventional retrieval agents, SCI combines graph retrieval with corpus interaction tools to achieve greater interaction efficiency. Although NeMo Agent uses the fewest turns across all three backbones, it makes more tool calls per turn and incurs higher total generation costs. SCI achieves higher accuracy and lower generation costs with fewer tool calls, demonstrating its efficiency in organizing retrieval and evidence exploration.
End-to-end execution times are reported in Appendix~\ref{app:execution-time}.

\begin{table*}[t]
\caption{Main results on BrowseComp-Plus with a 100K-document corpus. Turns and Tools are averaged per question. Cost is the total generation cost in USD for all 830 queries. Bold indicates the best values.}
\label{tab:bcp}
\centering
\small
\definecolor{bcpgroup}{HTML}{F5EDDF}
\definecolor{bcpsci}{HTML}{E3F2E9}
\setlength{\tabcolsep}{2pt}
\renewcommand{\arraystretch}{1.15}
\begin{tabular*}{\textwidth}{@{\extracolsep{\fill}}lcccc@{\hspace{18pt}}cccc@{\hspace{18pt}}cccc@{}}
\toprule
& \multicolumn{4}{c}{\textbf{GPT-6-Luna}} & \multicolumn{4}{c}{\textbf{DeepSeek-V4.1-Flash}} & \multicolumn{4}{c}{\textbf{Qwen3.8-Flash}} \\
\cmidrule(l{4pt}r{18pt}){2-5}\cmidrule(l{4pt}r{18pt}){6-9}\cmidrule(l{4pt}){10-13}
Method & Acc.$\uparrow$ & Turns$\downarrow$ & Tools$\downarrow$ & Total Cost$\downarrow$ & Acc.$\uparrow$ & Turns$\downarrow$ & Tools$\downarrow$ & Total Cost$\downarrow$ & Acc.$\uparrow$ & Turns$\downarrow$ & Tools$\downarrow$ & Total Cost$\downarrow$ \\
\midrule
\rowcolor{bcpgroup}
\multicolumn{13}{l}{\textit{Conventional retrieval agents}} \\
Retrieval-Agent & 80.00 & 16.43 & 35.97 & \$33.15 & 87.95 & 20.47 & 36.57 & \$26.22 & 82.41 & 24.54 & 52.64 & \$133.87 \\
NeMo Agent & 72.05 & \textbf{8.09} & 27.64 & \$23.06 & 80.12 & \textbf{10.47} & 30.69 & \$25.72 & 76.87 & \textbf{15.77} & 31.86 & \$56.86 \\
\midrule
\rowcolor{bcpgroup}
\multicolumn{13}{l}{\textit{Corpus interaction agents}} \\
DCI & 78.07 & 19.94 & 29.85 & \$52.83 & 83.86 & 23.43 & 34.79 & \$40.77 & 84.58 & 33.51 & 37.27 & \$61.46 \\
DR-DCI & 77.35 & 19.78 & 23.98 & \$13.58 & 85.78 & 21.96 & 32.54 & \$36.76 & 86.99 & 25.85 & 34.42 & \$66.76 \\
RISE & 74.70 & 17.17 & 16.76 & \$16.79 & 92.65 & 21.58 & 24.75 & \$15.77 & 90.12 & 22.50 & 27.17 & \$35.48 \\
RARG++ & 88.92 & 14.65 & 18.97 & \$14.73 & 93.86 & 31.88 & 39.29 & \$19.08 & 86.14 & 49.03 & 57.86 & \$61.47 \\
AtlasNav & 79.40 & 11.78 & 30.35 & \$18.21 & 93.61 & 15.98 & 35.89 & \$21.61 & 91.20 & 16.46 & 26.71 & \$33.21 \\
\textbf{SCI} & \textbf{91.08} & 11.22 & \textbf{16.14} & \textbf{\$11.82} & \textbf{95.30} & 16.68 & \textbf{18.07} & \textbf{\$11.72} & \textbf{94.94} & 17.08 & \textbf{18.32} & \textbf{\$30.41} \\
\bottomrule
\end{tabular*}
\end{table*}

\textbf{TREC-COVID document ranking.}
As shown in Table~\ref{tab:treccovid}, SCI achieves 81.77\% nDCG@10. This exceeds RARG++, the strongest baseline, by 5.80 percentage points. Compared with RARG++, SCI reduces average turns, tool calls, and generation costs by 56.3\%, 33.6\%, and 48.3\%, respectively. RISE-BM25 uses fewer tool calls but has substantially lower ranking quality. 
These results suggest that CEP's ranking helps prioritize more relevant documents within the agent's observation window, improving final ranking quality. 

\begin{table}[h]
\caption{Main results on TREC-COVID with GPT-6-Luna. Turns and Tools are averaged per query. Cost is the total generation cost for 50 queries. Best values are bold.}
\label{tab:treccovid}
\centering
\begin{tabular}{@{}lcccc@{}}
\toprule
Method & nDCG@10$\uparrow$ & Turns$\downarrow$ & Tools$\downarrow$ & Total Cost$\downarrow$ \\
\midrule
DCI &68.85&10.18&56.00&\$1.11\\
DR-DCI &67.17&6.20&23.20&\$0.16\\
RISE-BM25 &58.42&9.14&\textbf{10.72}&\$0.19\\
RARG++ &75.97&11.98&20.66&\$0.29\\
AtlasNav &57.79&7.00&25.06&\$0.26\\
SCI &\textbf{81.77}&\textbf{5.24}&13.72&\textbf{\$0.15}\\
\bottomrule
\end{tabular}
\end{table}

\begin{figure}[h]
\centering
\includegraphics[width=\columnwidth]{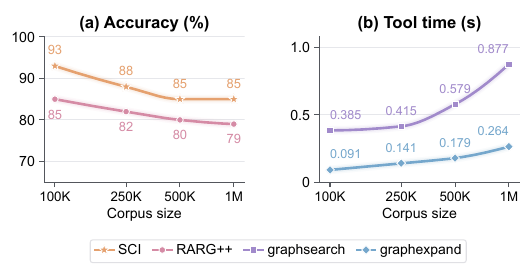}
\caption{Corpus scaling results. (a) Accuracy of SCI and RARG++. (b) Mean tool time of \textbf{\texttt{graphsearch}} and \textbf{\texttt{graphexpand}}.}
\label{fig:corpus-scaling}
\Description{Accuracy of SCI and RARG++ and mean execution times of graphsearch and graphexpand as the corpus grows from 100K to 1M documents. SCI maintains higher accuracy across all tested sizes. At 1M documents, graphsearch and graphexpand take 0.877 and 0.264 seconds on average, respectively.}
\end{figure}

\textbf{Corpus Scaling and Tool Efficiency.}
We evaluate corpus scaling and tool efficiency on a subset of 100 BrowseComp-Plus queries. To evaluate SCI in larger search spaces, we sample 900K documents from FineWeb-Edu~\cite{finewebedu} and progressively expand the BrowseComp-Plus corpus from 100K to 250K, 500K, and 1M documents. As shown in Figure~\ref{fig:corpus-scaling}(a), accuracy declines for both SCI and RARG++ as the corpus grows, but SCI consistently achieves higher accuracy, reaching 85\% at 1M documents. Figure~\ref{fig:corpus-scaling}(b) shows that after a roughly tenfold corpus expansion, mean execution times for \textbf{\texttt{graphsearch}} and \textbf{\texttt{graphexpand}} are 0.877 and 0.264 seconds, respectively. These tool latencies remain acceptable for interactive retrieval, supporting continued retrieval and expansion as the corpus grows.

\subsection{Ablation Study}\label{sec:ablation}
We conduct ablations on BrowseComp-Plus to examine retrieval algorithms, CEP components, and the structured map. The first two ablation experiments evaluate initial retrieval performance using SCI's initial retrieval requests for all 830 queries. Structured map ablations evaluate the full agent process on a subset of 100 queries.

\textbf{Retrieval algorithms.}
We replace CEP with three conventional retrieval methods (BM25, Dense, and Hybrid) and two graph retrieval methods (PPR and Prize-Collecting Steiner Tree (PCST)~\cite{g-retriever}). Dense uses the same Qwen3-Embedding-0.6B model as SCI, and Hybrid uses CEP's semantic and lexical fusion. PPR and PCST share CEP's document graph and starting weights. 
We set per-query retrieval budgets of 100, 500, and 1,000 documents and compute gold recall in the initial workspace. 
For fair comparison, we collect candidates and their path documents in score order, retaining only \(B\) unique documents per query.
We also compute gold recall for the top 10 preview documents per query to measure evidence coverage under a limited display budget.

\begin{table}[h]
\caption{Retrieval algorithm ablation on BrowseComp-Plus. Gold recall (\%) is measured after initial retrieval. PCST returns a selected subtree rather than a ranked list, so Top10 preview recall is not reported.}
\label{tab:retrieval}
\centering
\small
\setlength{\tabcolsep}{3pt}
\setbox0=\hbox{Retrieval budget per query}
\edef\budgetcolumnwidth{\the\dimexpr(\wd0-4\tabcolsep)/3\relax}
\begin{tabular}{@{}cl*{3}{>{\centering\arraybackslash}p{\budgetcolumnwidth}}c@{}}
\toprule
& & \multicolumn{3}{c}{Retrieval budget per query} & Top10 \\
Type & Method & 100 & 500 & 1,000 & preview \\
\midrule
\multirow{3}{*}{\shortstack{Conventional\\retrieval}} & BM25 & 34.74 & 50.80 & 58.31 & 19.85 \\
& Dense & 45.14 & 67.76 & 77.16 & 20.83 \\
& Hybrid & 41.92 & 62.02 & 71.28 & 20.33 \\
\midrule
\multirow{3}{*}{\shortstack{Graph\\retrieval}} & PPR & 45.83 & 65.05 & 74.12 & 21.88 \\
& PCST & 40.71 & 60.35 & 68.42 & -- \\
& CEP & \textbf{50.59} & \textbf{71.68} & \textbf{78.15} & \textbf{26.50} \\
\bottomrule
\end{tabular}
\end{table}

As shown in Table~\ref{tab:retrieval}, CEP achieves the highest gold recall across all retrieval budgets and Top10 previews. At a budget of 1,000, CEP achieves 78.15\% recall, outperforming Dense and PPR. At budgets of 100 and 500, it reaches 50.59\% and 71.68\%, exceeding the strongest respective baselines by 4.76 and 3.92 percentage points. These results show that CEP retrieves and presents more gold documents under limited budgets, improving evidence coverage in the workspace and the agent's observation window.

\begin{figure}[htbp]
\centering
\includegraphics[width=\columnwidth]{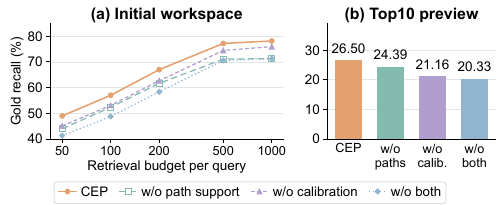}
\caption{CEP component ablation. (a) Initial workspace gold recall across retrieval budgets. (b) Top10 preview gold recall.}
\label{fig:components}
\Description{Workspace recall curves and preview recall bars for full CEP, without path support, without background calibration, and without both.}
\end{figure}

\textbf{CEP components.}
We remove path support, background calibration, or both to assess their contributions to initial retrieval performance. Removing path support replaces strongest path support with starting weights while retaining the same background distribution. Removing calibration retains path propagation. All four configurations use the same queries and budgets. We measure workspace gold recall at per-query retrieval budgets from 50 to 1,000 documents and Top10 preview recall.

As shown in Figure~\ref{fig:components}(a), full CEP achieves the highest workspace recall across all retrieval budgets. Removing either path support or background calibration reduces evidence coverage. Figure~\ref{fig:components}(b) shows that both components also improve Top10 preview recall, with background calibration having a larger effect. Together, these components help CEP retrieve more gold documents and present them within the agent's limited observation window. Section~\ref{sec:mechanism} further analyzes how background calibration mitigates ranking bias induced by graph structure.

\textbf{Structured map.}
To assess how the structured map and local expansion support evidence exploration, we remove each component separately and both together. Without map structure, we replace it with a flat list of document titles and previews, retaining CEP retrieval and \textbf{\texttt{graphexpand}}. When local expansion is disabled, we retain the structured map.

\begin{table}[t]
\caption{Structured map ablation on BrowseComp-Plus with GPT-6-Luna. Turns and tool calls are averaged per question. Bold indicates the best values.}
\label{tab:interaction}
\centering
\small
\setlength{\tabcolsep}{1pt}
\begin{tabular*}{\columnwidth}{@{\extracolsep{\fill}}lccccccc@{}}
\toprule
Configuration & Acc.$\uparrow$ & Turns$\downarrow$ & Tools$\downarrow$ & graphsearch & graphexpand & read & bash \\
\midrule
SCI & \textbf{93} & 14.15 & \textbf{18.61} & 5.15 & 0.88 & 6.68 & 5.90 \\
w/o map & 83 & \textbf{12.52} & 19.35 & 5.47 & 0.70 & 7.57 & 5.61 \\
w/o expansion & 83 & 13.08 & 21.40 & 4.46 & -- & 7.93 & 9.01 \\
w/o both & 80 & 12.66 & 21.95 & 4.74 & -- & 8.39 & 8.82 \\
\bottomrule
\end{tabular*}
\end{table}

As shown in Table~\ref{tab:interaction}, removing map structure or disabling local expansion reduces accuracy from 93\% to 83\%, while removing both reduces it to 80\%. Without map structure, the agent makes more global search and document reading calls. Disabling local expansion substantially increases \textbf{\texttt{read}} and \textbf{\texttt{bash}} calls, indicating greater reliance on reading individual documents and keyword searches. Although these variants use fewer turns on average, they make more total tool calls per query. These results indicate that removing source path guidance or the ability to expand from anchors makes it harder for the agent to discover and follow key evidence.

\begin{figure}[t]
\centering
\includegraphics[width=\linewidth]{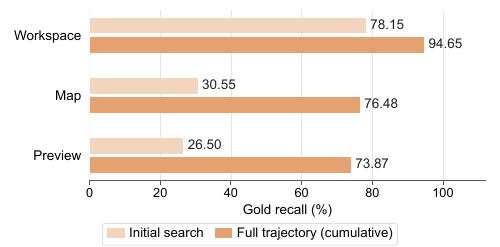}
\caption{Evolution of evidence coverage. Coverage after the initial search versus cumulative coverage over the full agent trajectory.}
\label{fig:coverage}
\Description{First-search and cumulative gold recall for workspace documents, map sources, and preview sources.}
\end{figure}

\subsection{Agent Behavior Analysis}
We analyze SCI's full trajectories for all 830 BrowseComp-Plus queries to examine evidence coverage during exploration and the contribution of local expansion.

\textbf{Evolution of evidence coverage.}
We measure source coverage at three levels: the workspace, map, and previews. The workspace contains documents whose full text has been collected. The map includes source documents and paths. Previews show the top 10 ranked source documents. As shown in Figure~\ref{fig:coverage}, from the first search to the end of exploration, workspace gold recall rises from 78.15\% to 94.65\%, cumulative map coverage from 30.55\% to 76.48\%, and cumulative preview coverage from 26.50\% to 73.87\%. During exploration, the agent acquires more gold documents and accesses their structural information and text previews via tool interactions.

\textbf{Evidence added by local expansion.}
On 511 of the 830 queries, the agent calls \textbf{\texttt{graphexpand}}. As shown in Table~\ref{tab:expansion}, these calls add 67 gold documents to the workspace and bring 372 previously collected but undisplayed gold documents into the map for the first time. Of these, 37 new documents and 212 existing documents are subsequently read using \textbf{\texttt{read}} or \textbf{\texttt{bash}}. These results show that local expansion both adds gold documents and exposes previously undisplayed gold documents already in the workspace to the agent, supporting subsequent reading. 
Appendix~\ref{app:expansion-cases} presents cases of local expansion with the returned maps.

\begin{table}[h]
\caption{Gold documents newly added to the workspace or first shown in the map from the existing workspace by local expansion. The last two columns count documents later accessed via \texttt{read} and via \texttt{read} or \texttt{bash}, respectively.}
\label{tab:expansion}
\centering
\small
\setlength{\tabcolsep}{2pt}
\begin{tabularx}{\columnwidth}{@{}l*{3}{>{\centering\arraybackslash}X}@{}}
\toprule
Expansion outcome & Documents & Read & \mbox{Read or Bash} \\
\midrule
Newly acquired & 67 & 32 & 37 \\
Newly shown in map & 372 & 180 & 212 \\
\bottomrule
\end{tabularx}
\end{table}

\subsection{Mechanism Analysis}\label{sec:mechanism}
We examine how CEP mitigates ranking bias caused by graph connectivity using SCI's initial retrieval requests for all 830 queries in BrowseComp-Plus. Our analysis covers preferences for highly connected nodes, gold document rankings, and changes in document rankings after adding distractor connections.

\textbf{Preference for highly connected nodes.}
Using the same document graph and starting weights, we compare the mean degree and proportion of highly connected nodes among the top 10, 100, and 1,000 documents returned by PPR and CEP. We define highly connected nodes as those with degrees at or above the 90th percentile of the full graph.
\begin{table}[htbp]
\caption{Mean degree and proportion of highly connected nodes among documents retrieved by PPR and CEP.}
\label{tab:structural-preference}
\centering
\small
\setlength{\tabcolsep}{4pt}
\begin{tabularx}{\columnwidth}{@{}l*{4}{>{\centering\arraybackslash}X}@{}}
\toprule
& \multicolumn{2}{c}{Mean degree} & \multicolumn{2}{c}{Highly connected (\%)} \\
\cmidrule(lr){2-3}\cmidrule(lr){4-5}
Top-$k$ & PPR & CEP & PPR & CEP \\
\midrule
10 & 182.29 & 66.56 & 56.76 & 33.55 \\
100 & 220.63 & 57.29 & 56.13 & 29.36 \\
1,000 & 171.00 & 61.30 & 50.26 & 30.68 \\
\bottomrule
\end{tabularx}
\end{table}

As shown in Table~\ref{tab:structural-preference}, CEP retrieves documents with lower mean degrees and smaller proportions of highly connected nodes than PPR at all three cutoffs. This consistent pattern indicates that CEP relies less on graph connectivity when ranking candidates.

\begin{figure}[h]
\centering
\includegraphics[width=\columnwidth]{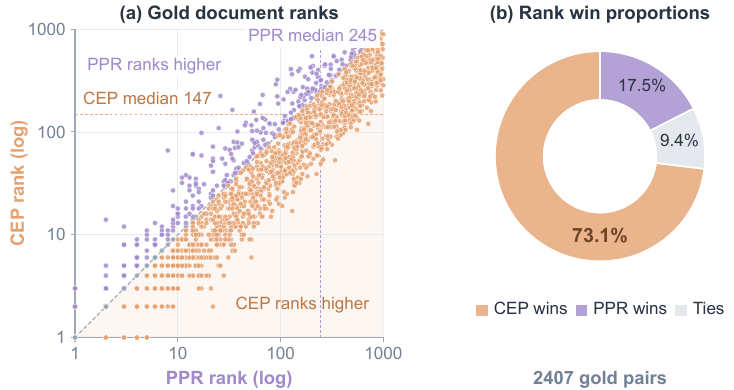}
\caption{Gold document rankings under PPR and CEP. (a) Rank pairs within the top 1,000 under both methods, with medians computed over all 2,407 rank pairs. (b) Proportions of rank wins and ties across all rank pairs.}
\label{fig:gold-rank-pairs}
\end{figure}

\textbf{Gold document rankings.}
For each question's gold documents, we pair their best ranks under PPR and CEP across the initial retrieval subqueries. As shown in Figure~\ref{fig:gold-rank-pairs}(a), CEP improves the median gold document rank from 245 under PPR to 147. Across all 2,407 rank pairs, CEP ranks the gold document higher than PPR in 73.1\% of cases (Figure~\ref{fig:gold-rank-pairs}(b)). Together with Table~\ref{tab:structural-preference}, these results show that CEP reduces preference for highly connected nodes while improving gold document rankings.

\textbf{Controlled edge addition.}
We test whether background calibration suppresses ranking advantages from additional connections. For each query, we select one document outside the gold set as a distractor target. The target must rank between 11 and 100 under both PPR and full CEP for the same subquery. Keeping document text, queries, and starting weights fixed, we randomly connect the target to documents outside the gold set that have lower starting weights and are not already connected to it. We repeat the edge addition experiment 10 times per target and measure its mean rank and Top10 entry rate for the selected subquery.

\begin{table}[htbp]
\caption{Mean rank and Top10 entry rate of distractor documents before and after controlled edge addition. Higher mean ranks and lower Top10 entry rates indicate stronger suppression of distractors.}
\label{tab:distractor-edges}
\centering
\small
\setlength{\tabcolsep}{3pt}
\begin{tabularx}{\columnwidth}{@{}l*{2}{>{\centering\arraybackslash}X}@{}}
\toprule
Method & \shortstack{Mean distractor rank\\Before $\rightarrow$ After} & \shortstack{Top10 entry (\%)\\Before $\rightarrow$ After} \\
\midrule
PPR & $50.67 \rightarrow 24.96$ & $0.00 \rightarrow 15.22$ \\
CEP w/o calibration & $42.21 \rightarrow 42.21$ & $0.84 \rightarrow 0.84$ \\
Full CEP & $45.79 \rightarrow 75.87$ & $0.00 \rightarrow 0.00$ \\
\bottomrule
\end{tabularx}
\end{table}

As shown in Table~\ref{tab:distractor-edges}, adding connections moves the mean rank of distractor documents under PPR from 50.67 to 24.96 and increases their Top10 entry rate from 0 to 15.22\%. In contrast, full CEP moves their mean rank from 45.79 to 75.87 while keeping the Top10 entry rate at 0, whereas the mean rank remains nearly unchanged without background calibration. These results show that background calibration suppresses ranking advantages arising from graph connectivity, making it harder for distractors to rank higher solely because they have more connections.

\section{Conclusion}
We present Contrastive Evidence Paths (CEP) and Structured Corpus Interaction (SCI) to address candidate ranking bias induced by graph connectivity and the lack of source paths for continued evidence exploration. CEP combines query-conditioned strongest path support with background visitation calibration to prioritize candidate documents and retain source paths. SCI uses these paths through a structured map and anchor expansion for source verification and continued exploration. Experiments on BrowseComp-Plus and TREC-COVID show that SCI improves answer accuracy, document ranking quality, and interaction efficiency while reducing generation costs. Further analysis shows that CEP mitigates connectivity-induced ranking bias and improves gold document rankings and evidence coverage. Anchor expansion also brings previously undisplayed gold documents in the workspace into the agent's observation window. We hope this work inspires future research to better integrate graph structures and graph retrieval into how agents solve problems, supporting continued evidence exploration within a limited observation window.



\clearpage 

\bibliographystyle{ACM-Reference-Format}
\bibliography{references}


\appendix

\section{Equivalence of Edgewise and Endpoint Calibration}\label{app:equivalence}
The main text first computes strongest path support and then calibrates candidate scores using the background visitation probability of each endpoint. Our implementation incorporates this calibration into initialization and edgewise propagation. We show that both formulations yield identical candidate scores and sets of optimal paths under the same path set and hop limit.

Uniform restart ensures $b_Q(v)>0$. Define $h_Q(v)=b_Q(v)^{-1/2}$. Initialize the starting weights as $p_i(v)h_Q(v)$ and define the edgewise scoring factors
\begin{equation}
\widetilde K_Q(u,v)=\lambda K_Q(u,v)\frac{h_Q(v)}{h_Q(u)}.
\label{eq:13}
\end{equation}

For any path $\pi=(v_0,\ldots,v_\ell)$, let its uncalibrated score be
\begin{equation*}
w_i(\pi)=p_i(v_0)\lambda^\ell
\prod_{j=1}^{\ell}K_Q(v_{j-1},v_j).
\end{equation*}

Its score under edgewise calibration is
\begin{equation}
\begin{aligned}
\widetilde w_i(\pi)
&=p_i(v_0)h_Q(v_0)
  \prod_{j=1}^{\ell}\widetilde K_Q(v_{j-1},v_j)\\
&=w_i(\pi)h_Q(v_0)
  \prod_{j=1}^{\ell}\frac{h_Q(v_j)}{h_Q(v_{j-1})}\\
&=h_Q(v_\ell)w_i(\pi).
\end{aligned}
\label{eq:14}
\end{equation}

The intermediate background factors cancel, leaving only the endpoint factor. This identity also holds for zero-hop paths.

Let $\Pi_H(v)$ denote the set of paths ending at $v$ with at most $H$ hops. Since $h_Q(v)$ is a positive constant independent of the particular path,
\begin{equation}
\begin{aligned}
\max_{\pi\in\Pi_H(v)}\widetilde w_i(\pi)
&=h_Q(v)\max_{\pi\in\Pi_H(v)}w_i(\pi)\\
&=\frac{M_i(v)}{\sqrt{b_Q(v)}}=S_i(v).
\end{aligned}
\label{eq:app-score-equivalence}
\end{equation}

Thus, edgewise calibration yields the same candidate scores as endpoint calibration in Eq.~\eqref{eq:8}. For a fixed endpoint, all paths receive the same multiplicative factor, preserving the set of optimal paths. Different endpoint factors can change the ranking across candidates.

This relation also holds at every dynamic programming iteration. Initialize $\widetilde m_i^{(0)}(v)=p_i(v)h_Q(v)$ and update
\begin{equation}
\begin{aligned}
\widetilde m_i^{(t+1)}(v)=\max\Bigl\{&p_i(v)h_Q(v),\\
&\max_{\mathclap{u:P(u,v)>0}}\widetilde K_Q(u,v)
\widetilde m_i^{(t)}(u)\Bigr\}.
\end{aligned}
\label{eq:app-edgewise-recurrence}
\end{equation}

Following the main text, the maximum over an empty predecessor set is zero. Induction gives
\begin{equation*}
\widetilde m_i^{(t)}(v)=h_Q(v)m_i^{(t)}(v).
\end{equation*}

After the same $H$ iterations, the edgewise recurrence therefore directly yields $S_i(v)$ and preserves the optimal backtracking choices. When multiple paths tie, identical tie-breaking rules recover the same path.

Finally, $\widetilde K_Q$ contains scoring factors, not transition probabilities, and individual factors may exceed $1$. For any nonempty cycle $C$, however, the background factors cancel:
\begin{equation}
\prod_{(u,v)\in C}\widetilde K_Q(u,v)
=\lambda^{|C|}\prod_{(u,v)\in C}K_Q(u,v)<1,
\label{eq:app-cycle-product}
\end{equation}
where $|C|$ is the number of edges in the cycle, $0<\lambda<1$, and $0\leq K_Q(u,v)\leq1$. Traversing a cycle cannot increase the path score, and removing it cannot reduce the calibrated score for the same endpoint.

\section{Implementation and Experimental Details}
\subsection{Experimental Details}\label{app:experiments}
This section details the model configurations, graph construction and retrieval parameters, execution and evaluation settings, and cost accounting.

\textbf{Model configuration.}
SCI uses high reasoning effort with GPT-6-Luna and DeepSeek-V4.1-Flash. Qwen3.8-Flash and Qwen3.8-27B~\cite{qwen38}, the latter used in Appendix~C.1, both use xhigh reasoning effort. Our local GPU environment consists of eight NVIDIA A100 80GB GPUs.

\textbf{Graph construction and retrieval parameters.}
For entity recognition, we use GLiNER2.5-small with a confidence threshold of 0.5 to extract eight entity types: person, organization, location, creative work, event, award, profession or role, and disease or medical condition. For semantic retrieval, we split documents into segments using the Qwen3-Embedding-0.6B tokenizer, limiting the full encoded input of each segment to 1,024 tokens with an overlap of 128 tokens between adjacent text segments. To reduce storage costs, we retain the first 256 dimensions of Qwen3-Embedding-0.6B output vectors and apply $L_2$ normalization. Table~\ref{tab:config} summarizes the retrieval parameters used in SCI.

\begin{table}[t]
\caption{SCI retrieval parameters.}
\label{tab:config}
\begin{tabularx}{\columnwidth}{@{}Xl@{}}
\toprule
Parameter & Setting \\
\midrule
Entity document frequency limit $\tau$ & 4 \\
Rank offset $k$ & 60 \\
Retrieval starting points per query $R$ & 1000 \\
Background restart probability $\alpha$ & 0.5 \\
Background estimation iterations $T_b$ & 24 \\
Hop discount $\lambda$ & 0.5 \\
Maximum hops for \textbf{\texttt{graphsearch}} $H$ & 3 \\
Candidate budget per query $B$ & 1000 \\
Maximum hops for \textbf{\texttt{graphexpand}} & 2 \\
Local expansion beam width & 64 \\
Local expansion candidate limit & 100 \\
\bottomrule
\end{tabularx}
\end{table}

\textbf{Execution and evaluation.}
All agent methods use at most 100 turns and a time budget of 3,600 seconds per query. Runtime includes the time spent on successful model requests and tool execution.
On BrowseComp-Plus, agents generate final answers through retrieval and evidence exploration. We use DeepSeek-V4.1-Flash throughout to judge whether generated answers are semantically equivalent to the reference answers. Accuracy is the proportion of queries answered correctly.
On TREC-COVID, we use the document titles, text, and relevance judgments provided by BEIR. Agents return a ranked list of up to 10 distinct document IDs in descending order of relevance. We compute nDCG@10 using the relevance judgments and average it across all 50 queries.

\textbf{Cost accounting.}
Table~\ref{tab:pricing} summarizes the generation rates used for cost calculation.

\begin{table}[t]
\caption{Generation pricing in USD per million tokens. GPT-6-Luna tiers refer to input tokens per request. DeepSeek-V4.1-Flash uses off-peak rates.}
\label{tab:pricing}
\centering
\small
\setlength{\tabcolsep}{2pt}
\begin{tabular*}{\columnwidth}{@{\extracolsep{\fill}}lrrrr@{}}
\toprule
Model & Input & Cache hit & Cache write & Output \\
\midrule
GPT-6-Luna ($\leq$272K) & 0.10 & 0.01 & 0.125 & 0.50 \\
GPT-6-Luna ($> $272K) & 0.20 & 0.02 & 0.25 & 0.75 \\
DeepSeek-V4.1-Flash & 0.15 & 0.003 & -- & 0.60 \\
Qwen3.8-Flash & 0.15 & 0.016 & 0.20 & 0.47 \\
\bottomrule
\end{tabular*}
\end{table}

\subsection{Agent Implementation}\label{app:interaction}
We implement SCI using Pi Agent~\cite{piagent}, with a separate workspace and interaction history for each query.

\textbf{Agent loop and tools.}
Pi Agent manages model requests, tool execution, and conversation history. For corpus interaction, the agent uses four tools: \textbf{\texttt{read}}, \textbf{\texttt{bash}}, \textbf{\texttt{graphsearch}}, and \textbf{\texttt{graphexpand}}. The graph tools collect source documents and update the structured map, while \textbf{\texttt{read}} and \textbf{\texttt{bash}} operate on the acquired workspace.

\textbf{Context management.}
We use a tool history compaction strategy similar to that of DR-DCI~\cite{drdci}. When the input length reaches 200,000 tokens, we replace older tool result bodies with \texttt{[cleared]} in subsequent model requests. We retain at least the latest 12 tool results and preserve the complete tool call group at the retention boundary. System instructions, the query, agent messages, and tool call records remain unchanged.

\textbf{Previews and tool output.}
For each displayed document, we select the two indexed segments most semantically relevant to the corresponding retrieval or expansion query, then use BM25 scores to choose a continuous text window within these segments. If no query terms match, we use the beginning of the highest-ranked segment. Preview lengths are limited to 240 characters for \textbf{\texttt{graphsearch}} and 600 for \textbf{\texttt{graphexpand}}. Previews preserve the original text and include source paths and character offsets for subsequent reading. Identical previews shown earlier are replaced with a brief notice. For \textbf{\texttt{read}} and \textbf{\texttt{bash}}, we follow DR-DCI\textquotesingle{}s tool output handling~\cite{drdci}, limiting returned text to 10 KiB and providing continuation information for truncated output.

\section{Supplementary Comparisons}\label{app:supplementary}
We conduct two supplementary experiments on a subset of 100 BrowseComp-Plus queries, examining agent performance with smaller models and conventional graph-based RAG methods under different execution modes.

\subsection{Smaller Models and Specially Trained Agents}
To examine SCI's performance with a smaller model, we compare SCI with five corpus interaction baselines on a subset of 100 BrowseComp-Plus queries, using Qwen3.8-27B~\cite{qwen38} as the common backbone. We additionally include OpenResearcher-30B-A3B~\cite{OpenResearcher} and QUEST-35B-RL~\cite{QUEST}, specially trained deep research agents with similar total parameter counts.

\begin{table}[t]
\caption{Supplementary agent comparison on 100 BrowseComp-Plus queries.}
\label{tab:supp-agents}
\centering
\definecolor{bcpgroup}{HTML}{F5EDDF}
\definecolor{bcpsci}{HTML}{E3F2E9}
\small
\setlength{\tabcolsep}{4pt}
\begin{tabularx}{\columnwidth}{@{}l*{3}{>{\centering\arraybackslash}X}@{}}
\toprule
Method & Acc.$\uparrow$ & Turns$\downarrow$ & Tools$\downarrow$ \\
\midrule
\rowcolor{bcpgroup}
\multicolumn{4}{l}{\textit{Specially trained deep research agents}} \\
OpenResearcher-30B-A3B & 46.00 & 67.64 & 66.44 \\
QUEST-35B-RL & 48.00 & 56.14 & 53.79 \\
\midrule
\rowcolor{bcpgroup}
\multicolumn{4}{l}{\textit{Corpus interaction agents (Qwen3.8-27B)}} \\
DCI & 77.00 & 27.53 & 74.94 \\
DR-DCI & 71.00 & 27.60 & 70.26 \\
RISE & 77.00 & 27.69 & 34.55 \\
RARG++ & 83.00 & 39.91 & 49.24 \\
AtlasNav & 79.00 & 19.95 & 25.91 \\
SCI & \textbf{84.00} & \textbf{19.92} & \textbf{22.64} \\
\bottomrule
\end{tabularx}
\end{table}

As shown in Table~\ref{tab:supp-agents}, SCI achieves 84\% accuracy with Qwen3.8-27B and uses the fewest average turns and tool calls among all compared methods. Compared with specially trained deep research agents, methods equipped with corpus interaction tools generally achieve higher accuracy and interaction efficiency. Compared with RARG++, the baseline with the highest accuracy, SCI reduces turns and tool calls by 50.1\% and 54.0\%, respectively, while achieving higher accuracy. These results show that the advantages in accuracy and interaction efficiency observed in the main experiments extend to a smaller backbone model.

\subsection{Reference Comparison with Graph-Based RAG Methods}

We compare HippoRAG2~\cite{HippoRAG2}, PathRAG~\cite{PathRAG}, and LinearRAG~\cite{linearRAG}. HippoRAG2 and PathRAG rely on LLMs to extract entities and relations offline, making graph construction on the full BrowseComp-Plus corpus costly. With Qwen3.8-27B deployed on eight NVIDIA A100 80GB GPUs as two replicas with tensor parallelism (TP=4) each, graph extraction for both methods on the full corpus could take several weeks in total. Given this substantial construction time, we use a reduced corpus to compare these methods with SCI.

For a fixed set of 100 BrowseComp-Plus queries, we collect all 562 officially annotated gold documents and randomly sample 10 distractor documents per query from the official negatives. Deduplication yields 983 negative documents, which we combine with the gold documents into a shared retrieval corpus of 1,545 documents. The three graph RAG methods split documents into chunks of 1,200 tokens with an overlap of 100 tokens, yielding 11,612 chunks, and build their graph indexes separately. SCI uses the same corpus with its own graph construction and retrieval procedure.

All methods use GPT-6-Luna for online retrieval and answering. We evaluate the three graph RAG methods using both their native retrieval and answering pipelines and their retrieval modules integrated as tools into Pi Agent~\cite{piagent}. The latter uses the same budget as SCI, at most 100 turns and one hour per query. We also report SCI's results on the same 100 queries using the full corpus.

\begin{table}[t]
\caption{Reference comparison of graph-based RAG methods and SCI. Cost is the total generation cost in USD for 100 queries. For a fair cost comparison among agents, we only report costs for agent methods.}
\label{tab:supp-graph-rag}
\centering
\definecolor{bcpgroup}{HTML}{F5EDDF}
\definecolor{bcpsci}{HTML}{E3F2E9}
\small
\setlength{\tabcolsep}{2.5pt}
\begin{tabularx}{\columnwidth}{@{}>{\raggedright\arraybackslash}Xccccc@{}}
\toprule
Method & Corpus size & Acc.$\uparrow$ & Turns$\downarrow$ & Tools$\downarrow$ & Total Cost$\downarrow$ \\
\midrule
\rowcolor{bcpgroup}
\multicolumn{6}{l}{\textit{Native graph RAG pipelines}} \\
HippoRAG2 & 1,545 & 42.00 & --- & --- & --- \\
PathRAG & 1,545 & 28.00 & --- & --- & --- \\
LinearRAG & 1,545 & 28.00 & --- & --- & --- \\
\midrule
\rowcolor{bcpgroup}
\multicolumn{6}{l}{\textit{Graph retrieval tools within Pi Agent}} \\
HippoRAG2 + Pi Agent & 1,545 & 89.00 & 12.49 & 26.74 & \$2.43 \\
PathRAG + Pi Agent & 1,545 & 83.00 & 7.40 & 15.41 & \$5.13 \\
LinearRAG + Pi Agent & 1,545 & 85.00 & 11.43 & 26.98 & \$4.59 \\
\midrule
\rowcolor{bcpgroup}
\multicolumn{6}{l}{\textit{SCI across corpus sizes}} \\
SCI & 1,545 & \textbf{97.00} & \textbf{6.77} & \textbf{11.54} & \textbf{\$0.55} \\
SCI & 100,195 & 93.00 & 14.15 & 18.61 & \$1.96 \\
\bottomrule
\end{tabularx}
\end{table}

As shown in Table~\ref{tab:supp-graph-rag}, integrating the graph retrieval modules into Pi Agent increases the accuracies of HippoRAG2, PathRAG, and LinearRAG from 42\%, 28\%, and 28\% to 89\%, 83\%, and 85\%, respectively. On the same reduced corpus, SCI achieves 97\% accuracy with the fewest turns and tool calls and the lowest cost. 
These results show that SCI supports evidence exploration and answering more effectively under the same corpus and interaction budget.

When the retrieval scope expands to the full corpus, SCI still achieves 93\% accuracy at a total generation cost of \$1.96, lower than the costs of all three graph retrieval agents on the reduced corpus. 
This shows that SCI can sustain effective evidence exploration in a larger search space.

\begin{figure*}[t]
\centering
\includegraphics[width=\textwidth]{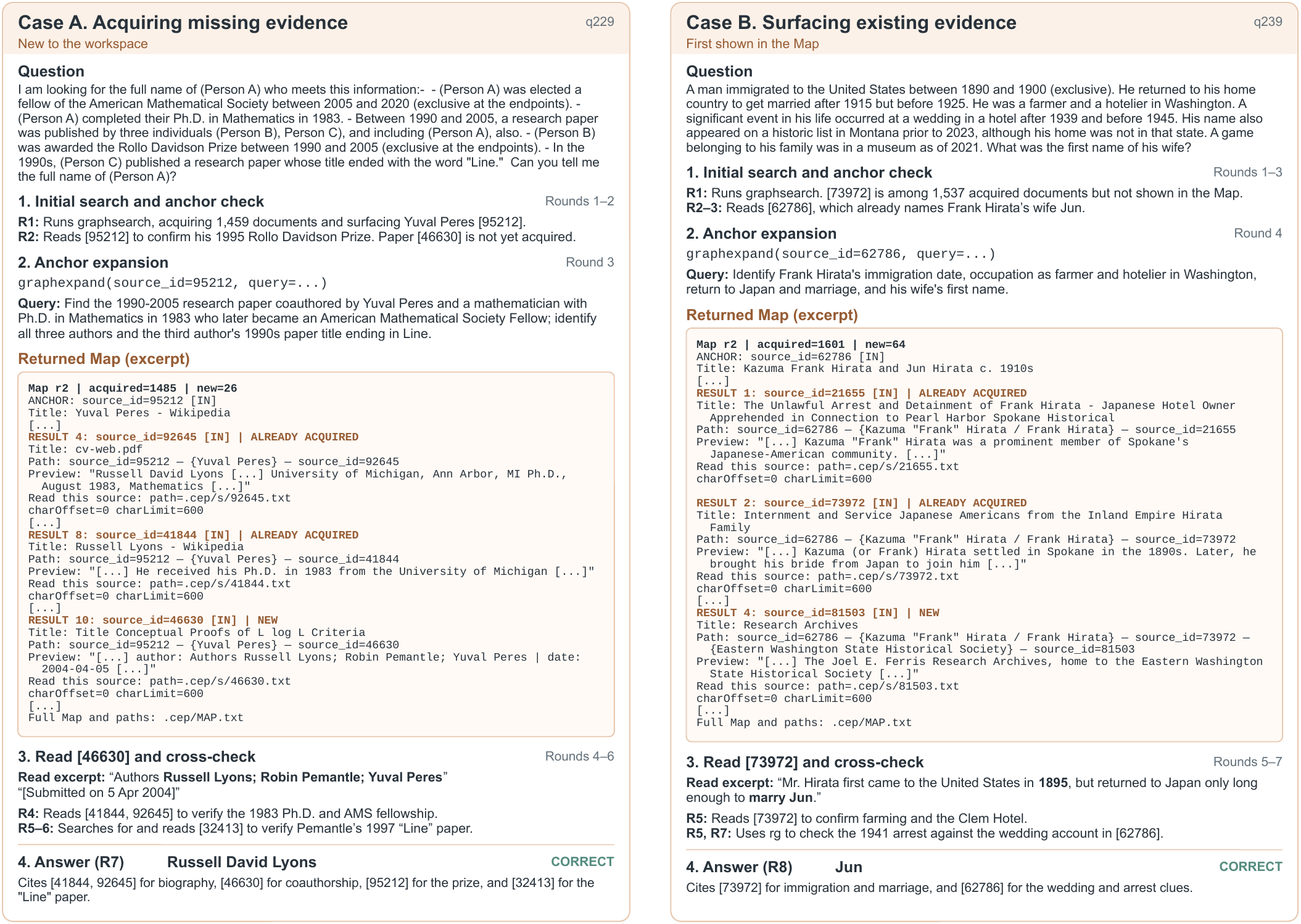}
\caption{Two local expansion cases on BrowseComp-Plus, with full questions, abbreviated tool calls, and returned map excerpts. Quotes reproduce source text. R denotes the interaction turn. Bracketed numbers are corpus document IDs.}
\label{fig:expansion-cases}
\Description{Two recorded local expansion trajectories. Case A retrieves a previously uncollected paper through Yuval Peres, then reads and cites it to verify a coauthorship clue supporting the answer Russell David Lyons. Case B first shows an already collected Hirata Family archive in the map, then reads and cites it to corroborate the answer Jun, whose name was already present in the anchor.}
\end{figure*}

\section{Limitations}
Our study focuses on evidence retrieval and continued exploration over a fixed document corpus. Maintaining the document graph as sources are added or revised is outside the scope of this work. Future work could extend SCI to evolving corpora through incremental graph updates. This includes updating entity associations and document connections while reusing existing computations, and keeping the graph consistent with the underlying corpus.


\section{Case Studies of Local Expansion}\label{app:expansion-cases}



Figure~\ref{fig:expansion-cases} illustrates two roles of local expansion. In Case A, expansion from the Yuval Peres document retrieves a previously uncollected paper, which the agent reads and cites to verify the coauthorship clue. In Case B, expansion from a Hirata photograph page surfaces an archive already in the workspace but not previously displayed, enabling verification of the immigration and marriage clues supporting the answer Jun.

\section{End-to-End Execution Time}
\label{app:execution-time}

\begin{table}[h]
\caption{Mean end-to-end execution time (s) per query on BrowseComp-Plus. Bold indicates the lowest values.}
\label{tab:wall-time}
\centering
\small
\setlength{\tabcolsep}{4pt}
\renewcommand{\arraystretch}{1.15}
\begin{tabular*}{\columnwidth}{@{\extracolsep{\fill}}lccc@{}}
\toprule
Method & \shortstack{GPT-6-\\Luna} & \shortstack{DeepSeek-\\V4.1-Flash} & \shortstack{Qwen3.8-\\Flash} \\
\midrule
Retrieval-Agent & 67.21\,s & 98.30\,s & 561.95\,s \\
NeMo Agent & 49.44\,s & 113.57\,s & 534.30\,s \\
DCI & 133.45\,s & 146.19\,s & 493.76\,s \\
DR-DCI & 62.04\,s & 164.89\,s & 652.18\,s \\
RISE & 59.57\,s & 111.83\,s & 419.95\,s \\
RARG++ & 172.86\,s & 232.50\,s & 960.02\,s \\
AtlasNav & 56.42\,s & 115.98\,s & 347.15\,s \\
\textbf{SCI} & \textbf{45.18\,s} & \textbf{62.11\,s} & \textbf{333.78\,s} \\
\bottomrule
\end{tabular*}
\end{table}

SCI achieves the lowest mean execution time across all three backbones. These results complement the tool usage and cost comparisons, showing that SCI's efficiency gains also translate into shorter execution times.

\end{document}